\documentclass[aps,twocolumn]{revtex4}
\usepackage[dvips]{graphicx}
\newcommand{\sect}[1]{\setcounter{equation}{0}\section{#1}}

\begin{document}
\title{RELATIVISTIC ENTROPY AND RELATED BOLTZMANN KINETICS}
\author{G. Kaniadakis}
\email{giorgio.kaniadakis@polito.it}
\affiliation{Dipartimento di Fisica, Politecnico di Torino, \\
Corso Duca degli Abruzzi 24, 10129 Torino, Italy}
\date{\today}
\begin {abstract}

It is well known that the particular form of the two-particle
correlation function, in the collisional integral of the classical
Boltzmman equation, fix univocally the entropy of the system, which
turn out to be the Boltzmann-Gibbs-Shannon entropy.

In the ordinary relativistic Boltzmann equation, some standard
generalizations, with respect its classical version, imposed by the
special relativity, are customarily performed.  The only ingredient
of the equation, which tacitly remains in its original classical
form, is the two-particle correlation function,  and this fact
imposes  that also the relativistic kinetics is governed by the
Boltzmann-Gibbs-Shannon entropy. Indeed the ordinary relativistic
Boltzmann equation admits as stationary stable distribution, the
exponential Juttner distribution.

Here, we show that the special relativity laws and the maximum
entropy principle, suggest a relativistic generalization also of the
two-particle correlation function and then of the entropy. The so
obtained, fully relativistic Boltzmann equation, obeys the H-theorem
and predicts a stationary stable distribution, presenting power-law
tails in the high energy region. The ensued relativistic kinetic
theory preserves the main features of the classical kinetics, which
recovers in the $c \rightarrow \infty$ limit.

\end {abstract}

%\pacs{PACS number(s): 05.90.+m, 05.20.-y, 51.10.+y, 03.30.+p}

\maketitle

\sect{Introduction}

In experimental high energy physics, the power law tailed
probability distribution functions \cite{EPJB}, have been observed
systematically (plasmas \cite{Hasegawa}, cosmic rays
\cite{Vasyliunas,Biermann}, particle production processes
\cite{Wilk,Walton} etc).

A mechanism, frequently used to explain the occurrence  of non
exponential distributions, is based on certain non-linear evolution
equations, mainly considered in the Fokker-Planck picture
\cite{Polynomial-clas,Polynomial-quant,Fractional,Quons,
Abeg,Frank1,Chavanis,Frank2,Curado,PhA01,PLA01,PRE02,PRE05}, but
recently also in the Boltzmann picture
\cite{PhA01,PRE02,PRE05,BiroK}. Clearly, the correctness of the
analytic expression of a given distribution, used to describe a
statistical system, is strongly related to the validity of its
generating mechanism.

The classical Boltzmann equation, due to the particular form of the
two-particle correlation function, in the collisional integral,
fix univocally the entropy of the system, which turn out to be
Boltzmann-Gibbs-Shannon entropy. The latter entropy imposes the
exponential form to the probability distribution function, emerging
as the stationary and stable solution of the equation.

In the ordinary relativistic Boltzmann equation \cite{DEGROOT}, some
standard generalizations, with respect its classical version, imposed
by the special relativity, are customarily performed. The only
ingredient of the equation, which tacitly remains in its original
classical form, is the two-particle correlation function, and this
fact imposes  that also the relativistic kinetics is governed by the
classical Boltzmann-Gibbs-Shannon entropy. Indeed the ordinary
relativistic Boltzmann equation admits as stationary stable
distribution, the exponential Juttner distribution.

In order to obtain a relativistic Boltzmann equation
admiting as stationary and stable solution a probability
distribution, different from the exponential one, the only
possibility we have, is to modify the expression of the classical
two-particle correlation function. As a consequence a modification
of the system entropy emerges, and this important fact suggests that
such modification must be confined possibly within the ordinary
physics (special relativity, maximum entropy principle) and without
invoking any additional assumption or postulate.

Main goal of the present paper is to show that it is possible to
obtain a fully relativistic Boltzmann equation admitting as
stationary and stable solution a distribution function presenting
power-law tails.

The new relativistic equation, can be obtained starting from the
ordinary relativistic Boltzmann equation \cite{DEGROOT}, by properly
generalizing the two-particle correlation function in the
collisional integral.

Such generalization, can be obtained in a self-consistent manner, by
employing (i) the special relativity laws and (ii) the maximum entropy
principle, and without invoking any extra principle
\cite{PhA01,PLA01,PRE02,PRE05}.

The new relativistic entropy and relevant probability distribution function
have very simple expressions, which result to be one-parameter
(light speed in dimensionless form) generalizations of the
corresponding classical concepts, just as happen for all the
physical quantities (energy, momentum etc) in special relativity.
Clearly, the relativistic entropy and the relativistic distribution
function, in the classical limit, reduce to the
Boltzmann-Gibbs-Shannon entropy and to the Maxwell-Boltzmann
distribution respectively.

Within the present theoretical framework, the power-law tails in the
distribution functions, in high energy physics, emerge as a
purely relativistic effect (those tails in the classical limit $c
\rightarrow \infty$ deform into exponential tails). In other words
the power-law tails represent the signature of the relativistic
nature of the system.

The paper is organized as follows:

In Sect. II we consider briefly the basic concepts of relativistic
dynamics by using dimensionless variables.

In Sect. III we consider the $\kappa$-differential calculus by
giving particular emphasis to its physical origin.

In Sect. IV  we present the main properties of the function
$\kappa$-exponential.

In Sect. V  we present the main properties of the function
$\kappa$-logarithm.

In Sect. VI  we introduce the velocity representation of the
functions $\kappa$-exponential and $\kappa$-logarithm.

In Sect. VII we consider Boltzmann relativistic kinetics and discuss
in detail the relativistic generalization of the classical
Boltzmann-Gibbs-Shannon entropy.

Finally in Sect. VIII we report some concluding remarks.

\sect{Relativistic dynamics with dimensionless variables}

In the present section, for simplicity of the exposition, we
consider briefly, the relativistic dynamics in one spatial
dimension. Starting from the velocity $v$, the momentum $p$ and the
total energy $E$ of a particle of rest mass $m$ we introduce the
dimensionless velocity $u$, momentum $q$, and total energy $\cal E$,
through
\begin{eqnarray}
\frac{v}{u}=\frac{p}{m q}=\sqrt{\frac{E}{m {\cal E}}}=\kappa c=v_* \
\ , \label{RII1}
\end{eqnarray}
being $c$ the light speed and $v_*<c$ a reference velocity. The
condition $-c<v<c$ implies that $-1/\kappa<u<1/\kappa$, so that
$1/\kappa$ represents the dimensionless light speed. Alternatively
$1/\kappa$ can be viewed as the refractive index of a medium in
which the light speed is $v_*$.  From the dispersion relation $E^2=
m^{2}c^{4}+ p^2c^2$ follows that $E\geq m c^2$ and consequently it
holds ${\cal E}\geq 1/\kappa^2$. Clearly the condition $-\infty < p
< +\infty$ implies $-\infty < q < +\infty$. For a particle at rest
it results $E(0)/{\cal E}(0)= m c^2\kappa^2$ and after posing
$E(0)=m\,c^2$ we obtain ${\cal E}(0)=1/\kappa^2$. Then we can
conclude that $1/\kappa^2$ represents the dimensionless rest energy
of the particle. It is important to note that the classical limit
$\lim_{c\rightarrow \infty}$ in dimensionless variable transforms
into the limit $\lim_{\kappa\rightarrow 0}$.

In relativistic dynamics any physical quantity or variable $h$, is
linked through an invertible function with the particle momentum i.e
$h=h(q)$, $q=q(h)$. The velocity $u$ and the total energy ${\cal E}$
are given by
\begin{eqnarray}
&&u(q)=\frac{q}{\sqrt{1+\kappa^2q ^2}} \ \ , \label{RII2} \\
&&{\cal E}(q)=\frac{1}{\kappa^2}\sqrt{1+\kappa^2q^2} \ \ ,
\label{RII3} \label{RII14}
\end{eqnarray}
while the kinetic energy ${\cal W}={\cal E}-1/\kappa^2 \geq 0$
assumes the form
\begin{eqnarray}
{\cal W}(q)=
\frac{1}{\kappa^2}\sqrt{1+\kappa^2q^2}-\frac{1}{\kappa^2} \ \ .
\label{RII4}
\end{eqnarray}

Clearly we can introduce further function $h=h(q)$ which define new
variables. For instance the Minkovski rapidity $\psi$, is defined
through $\beta=\tanh \psi$ with $\beta=v/c$. After introducing the
new rapidity variable $\rho$ through $\psi=\kappa \rho$, we can
express it in terms of dimensionless momentum as follows:
\begin{eqnarray}
\rho(q)=\frac{1}{\kappa}\,{\rm arcsinh} (\kappa q) \ \ .
\label{RII5}
\end{eqnarray}
The above definition of the rapidity variable is justified by the
fact that in the classical limit it results: $\rho=q=u=\sqrt{2{\cal
W}}$.

Let us consider in the inertial one-dimension spatial frame $\Sigma$, two
identical, non interacting, and free particles $A$ and $B$, of rest
mass $m$. We suppose that the particle $A$ moves toward right while
the particle $B$ moves toward left. The dimensionless variables for
the particle $A$ are given by $q_{\scriptscriptstyle A}$,
$v_{\scriptscriptstyle A}$, ${\cal E}_{\scriptscriptstyle A}$,
${\cal W}_{\scriptscriptstyle A}$ and $\rho_{\scriptscriptstyle A}$.
The absolute values of the momentum, velocity and rapidity  of the
particle $B$ are given by $q_{\scriptscriptstyle B}$,
$v_{\scriptscriptstyle B}$, $\rho_{\scriptscriptstyle B}$, while its
total energy and kinetic energy are given by ${\cal
E}_{\scriptscriptstyle B}$, ${\cal W}_{\scriptscriptstyle B}$
respectively.

In the rest frame ${\Sigma'}={\Sigma}_{\scriptscriptstyle B}$ of the
particle $B$, which moves toward left, with velocity
$v_{\scriptscriptstyle B}$ with respect $\Sigma$, the above
considered dimensionless variables, for the particle $B$ are given
by $q'_{\scriptscriptstyle B}=0$, $v'_{\scriptscriptstyle B}=0$,
${\cal E}'_{\scriptscriptstyle B}=1/\kappa^2$, ${\cal
W}'_{\scriptscriptstyle B}=0$ and $\rho'_{\scriptscriptstyle B}=0$.

According to the Lorentz transformations, the dynamic variables of
the particle $A$ i.e. $q'_{\scriptscriptstyle A}$,
$v'_{\scriptscriptstyle A}$, ${\cal E}'_{\scriptscriptstyle A}$,
${\cal W}'_{\scriptscriptstyle A}$ and $\rho'_{\scriptscriptstyle
A}$, in the frame ${\Sigma'}$, are direct compositions of the
corresponding variables of the particles $A$ and $B$  in the frame
${\Sigma}$. For instance the momentum composition law
\begin{eqnarray}
q'_{\scriptscriptstyle A}=q_{\scriptscriptstyle A\!}
\stackrel{\kappa}{\oplus} \,q_{\scriptscriptstyle B} \ ,
\label{RII6}
\end{eqnarray}
with
\begin{eqnarray}
q_{\scriptscriptstyle A\!} \stackrel{\kappa}{\oplus}
\,q_{\scriptscriptstyle B}=q_{\scriptscriptstyle A}\sqrt{1+\kappa^2
q_{\scriptscriptstyle B}^2}+ q_{\scriptscriptstyle
B}\sqrt{1+\kappa^2{q}_{\scriptscriptstyle A}^2} \ , \label{RII7}
\end{eqnarray}
follows directly from the Lorentz transformation of the four-vector
energy-momentum.

Regarding the composition law  of an arbitrary variable $h(q)$,
after posing $h_A=h(q_A)$, $h_B=h(q_B)$ and $h'_A=h(q'_A)$, one
obtains
\begin{eqnarray}
h'_A=h_{_A} \!\stackrel{\kappa}{\oplus}\!h_{_B} \ \ , \label{RII8}
\end{eqnarray}
with
\begin{eqnarray}
h_{_A} \!\stackrel{\kappa}{\oplus}\!h_{_B}=h\,(\,q_{_A}\!
\stackrel{\kappa}{\oplus}\!q_{_B}) \ \ , \label{RII9}
\end{eqnarray}
so that the composition law of any variable can be deduced from the
momentum composition law.

In particular starting from the general composition law defined in
Eq. (\ref{RII9}) and after taking into account Eqs.
(\ref{RII2})-(\ref{RII5}) we can obtain the composition laws for the
velocity, total energy, kinetic energy and rapidity as follows
\begin{eqnarray}
&&u'_{\scriptscriptstyle A}=u_{\scriptscriptstyle A\!}
\stackrel{\kappa}{\oplus} \,u_{\scriptscriptstyle B} \ ,
\label{RII10} \\
&&{\cal E}'_{\scriptscriptstyle A}={\cal E}_{\scriptscriptstyle A\!}
\stackrel{\kappa}{\oplus} \,{\cal E}_{\scriptscriptstyle B} \ ,
\label{RII11} \\
&&{\cal W}'_{\scriptscriptstyle A}={\cal W}_{\scriptscriptstyle A\!}
\stackrel{\kappa}{\oplus} \,{\cal W}_{\scriptscriptstyle B} \ ,
\label{RII12} \\
&&{\rho}'_{\scriptscriptstyle A}={\rho}_{\scriptscriptstyle A\!}
\stackrel{\kappa}{\oplus} \,{\rho}_{\scriptscriptstyle B} \ ,
\label{RII13}
\end{eqnarray}
being
\begin{eqnarray}
&& \!\!\!\!\!\!\!\!u_{\scriptscriptstyle A\!}
\stackrel{\kappa}{\oplus} \,u_{\scriptscriptstyle
B}=\frac{u_{\scriptscriptstyle A}+u_{\scriptscriptstyle
B}}{1+\kappa^2u_{\scriptscriptstyle A}u_{\scriptscriptstyle B}} \ ,
\label{RII14} \\
&&\!\!\!\!\!\!\!\!{\cal E}_{\scriptscriptstyle A\!}
\stackrel{\kappa}{\oplus} \,{\cal E}_{\scriptscriptstyle
B}=\kappa^2{\cal E}_{\scriptscriptstyle A} {\cal
E}_{\scriptscriptstyle B} +\frac{1}{\kappa^2}
\sqrt{\left(\kappa^4{\cal {E}}_{\scriptscriptstyle
A}^2-1\right)\left(\kappa^4{\cal
E}_{\scriptscriptstyle B}^2-1\right)} , \ \ \ \ \ \ \\
\label{RII15}
&&\!\!\!\!\!\!\!\!{\cal W}_{\scriptscriptstyle A\!}
\stackrel{\kappa}{\oplus} \,{\cal W}_{\scriptscriptstyle B}= {\cal
W}_{\scriptscriptstyle \!A}\!+\!{\cal W}_{\scriptscriptstyle \!B}
\!+\!\kappa^2{\cal W}_{\scriptscriptstyle \!A}
{\cal W}_{\scriptscriptstyle \!B}  + \nonumber \\
&&\ \ \ \ \ \ \ \ \ \ \  + \, \sqrt{{\cal W}_{\scriptscriptstyle
\!A} {\cal W}_{\scriptscriptstyle \!B}\big(2\!+\!\kappa^2{\cal
W}_{\scriptscriptstyle \!A}\big)\big(2\!+\!\kappa^2{\cal
W}_{\scriptscriptstyle \!B}\big)} \ , \label{RII16} \\
&&\!\!\!\!\!\!\!\!{\rho}_{\scriptscriptstyle A\!}
\stackrel{\kappa}{\oplus} \,{\rho}_{\scriptscriptstyle
B}={\rho}_{\scriptscriptstyle A} + {\rho}_{\scriptscriptstyle B} \ \
. \label{RII17}
\end{eqnarray}

It is remarkable that the momentum composition law
$q_{\scriptscriptstyle A}\! \stackrel{\kappa}{\oplus}
q_{\scriptscriptstyle B}$, is a generalized sum having the
properties: i) it is associative, ii) it is commutative, iii) the
$0$ is its neutral element, iv) the opposite element of $q$ is $-q$.
Also the velocity composition law $u_{\scriptscriptstyle A}\!
\stackrel{\kappa}{\oplus} u_{\scriptscriptstyle B}$, results to be a
generalized sum, while the rapidity composition law
$\rho_{\scriptscriptstyle A}\! \stackrel{\kappa}{\oplus}
\rho_{\scriptscriptstyle B}$ is the ordinary sum.

Regarding the total energy composition law ${\cal
E}_{\scriptscriptstyle A}\! \stackrel{\kappa}{\oplus} {\cal
E}_{\scriptscriptstyle B}$ we note that i) it is associative, ii) it
is commutative, iii) it has as neutral element the quantity
$1/\kappa^2$, iv) it does not exist the opposite element of ${\cal
E}$. Due to the latter property, the total energy composition law is
not a generalized sum. Also the kinetic energy composition law
${\cal W}_{\scriptscriptstyle A}\! \stackrel{\kappa}{\oplus} {\cal
W}_{\scriptscriptstyle B}$ is not a generalized sum because it does
not exist the opposite element of  ${\cal W}$.

\sect{The $\kappa$-Differential Calculus}

\subsection{Lorentz invariant integration}

Let us consider the following integral in the three dimension
momentum space, within the classical physics framework
\begin{eqnarray}
I_{cl}=\int \frac{d^3p}{p_*^{\,\,3}} \,F \ \ , \label{RIII1}
\end{eqnarray}
being $F$ an arbitrary function and $p_*=mc\kappa$. In dimensionless
variables, the latter integral becomes
\begin{eqnarray}
{\cal I}_{0}=\int d^3q \,F \ \ . \label{RIII2}
\end{eqnarray}
Whenever $F$ depends only on $q=|{\bf q}|$, the above integral can
be reduced to the following one dimension integral
\begin{eqnarray}
{\cal I}_{0}=\int_0^{\infty} d q\,\,4\pi\, q^2 \,F \ \ .
\label{RIII3}
\end{eqnarray}

In the framework of a relativistic theory it is well known that the
three-dimension integral (\ref{RIII1}) must be replaced by the
four-dimension Lorentz invariant integral
\begin{eqnarray}
I_{rel}= \int \frac{d^4p}{p_*^{\,\,3}\,m c}
\,\,\,2\,\theta\left(p_0\right)\,\delta\left(p^{\mu}p_{\mu}-m^2c^2\right)
\, F \ \ .  \ \ \ \  \label{RIII4}
\end{eqnarray}
After introducing the dimensionless variables according to
$q^{\mu}=p^{\mu}/p_*=(q_0, {\bf q})=(\kappa {\cal E}, {\bf q})$, the
latter integral becomes
\begin{eqnarray}
{\cal I}_{\kappa}= \int d^4q \,\,2\,\theta\left(\kappa
q_0\right)\,\delta\left(\kappa^2q^{\mu}q_{\mu}-1\right) \, \,F \ \
\label{RIII5} \ \ ,
\end{eqnarray}
or alternatively
\begin{equation}
{\cal I}_{\kappa}=\int d^3 q \int d (\kappa^2 {\cal E})
\,\,2\,\theta\left(\kappa^2{\cal E}
\right)\,\delta\left(\kappa^4{\cal E}^2-\kappa^2 q^2-1\right) \, \,F
\ \ \label{RIII6}
\end{equation}
with $q=|{\bf q}|$.

In order to reduce the integral (\ref{RIII6}), in a three-dimension
integral, we introduce the new integration variable $Z=\kappa^4
{\cal E}^2$. After observing that $d(\kappa^2 {\cal
E})=dZ/2\sqrt{Z}$ and $\theta(\kappa^2 {\cal E})=\theta(Z)$ one
obtains
\begin{eqnarray}
{\cal I}_{\kappa}=\int d^3 q \int_0^{+\infty}  \frac{d Z}{2\sqrt{Z}}
\,2 \,\delta\left(Z-\kappa^2q^2-1\right) \, \,F \ \ \label{RIII7}
\end{eqnarray}
and finally
\begin{eqnarray}
{\cal I}_{\kappa}=\int \frac{d^3q}{\sqrt{1+\kappa^2q^2}} \,\,F \ \ .
\label{RIII8}
\end{eqnarray}
The latter three-dimension integral in dimensional variables becomes
\begin{eqnarray}
I_{rel}=\int \frac{d^3p}{p_*^{\,3}}\,\, \frac{m c}{p_0} \,\,F \ \
\label{RIII9}
\end{eqnarray}

We remark that in Eq. (\ref{RIII4}) the integration element $d^4p$,
is a scalar because the Jacobian of the Lorentz transformation is
equal to unity. Then $I_{rel}$ transforms as $F$. For this reason in
Eq. (\ref{RIII8}) the integration element $d^3 q/\sqrt{1+\kappa^2
q^2 }$ is a scalar.

Whenever $F$ depends only on $q=|{\bf q}|$, the integral
(\ref{RIII8}) can be reduced to the following one dimension integral
\begin{eqnarray}
{\cal I}_{\kappa}=\int_0^{\infty} \frac{d q}{\sqrt{1+\kappa^2 q^2
}}\,\,4\pi\, q^2 \,F \ \ , \label{RIII10}
\end{eqnarray}
which after introducing the $\kappa$-differential
\begin{eqnarray}
d_{\kappa}q = \frac{d q}{\sqrt{1+\kappa^2q^2}} \ \ , \label{RIII11}
\end{eqnarray}
assumes the form
\begin{eqnarray}
{\cal I}_{\kappa}=\int_0^{\infty}d_{\kappa}q \,\,4\pi\, q^2 \,F \ \
. \label{RIII12}
\end{eqnarray}
In the classical limit $\kappa \rightarrow 0$, the latter integral
reproduces the corresponding classical one given by Eq.
(\ref{RIII3}).

We focus now our attention to the classical and relativistic
expression of the one dimension integrals given by (\ref{RIII3}) and
(\ref{RIII12}) respectively. One immediately observes that the
relativistic integral is obtained directly from the classical one,
by making the substitution $d q \rightarrow d_{\kappa}q$.

The equivalence of the Lorentz invariant integration in
four-dimension $\int d^4p$, with the $\kappa$-integration in
one-dimension $\int_0^{\infty}d_{\kappa}q$,  according to
\begin{equation}
\int d^4p
\,\,\,\theta\left(p_0\right)\,\delta\left(p^{\mu}p_{\mu}-m^2c^2\right)
\, F (p\,) \, \, \propto \int_0^{\infty}d_{\kappa}q \,\,\, q^2 F(q)\
\ , \ \ \ \  \label{RIII13}
\end{equation}
permits to explain better the relativistic origin of the
$\kappa$-integration.

\subsection{Kinetic Energy and Work}

In classical mechanics the kinetic energy $W$ is defined as the work
produced by the external force ${\bf f}$
\begin{eqnarray}
W_{cl}(p)&&=\int_0^p {\bf f}\,d{\bf x} \ \ . \label{RIII14}
\end{eqnarray}
After taking into account the Newton law and using dimensionless
variables the above definition assumes the form
\begin{eqnarray}
{\cal W}_{0}(q)=\int_0^q  q \,\, dq \ \ , \label{RIII15}
\end{eqnarray}
and yields ${\cal W}_{0}(q)=q^2/2.$

In special relativity, by using the same definition, the kinetic
energy is given by
\begin{eqnarray}
W_{rel}(p)\!\!\!&&=\int_0^p {\bf f}\,d{\bf x}=\int_0^p \frac{d{\bf
p}}{dt}\,d{\bf
x}= \int_0^p {\bf v}\,d{\bf p}   \nonumber \\
&&= \int_0^p v\,d p = \int_0^p
\frac{p/m_{_0}}{\sqrt{1+p^2/m_{_0}^2c^2}} \,\,dp \ \ . \ \ \ \ \ \ \
\ \ \label{RIII16}
\end{eqnarray}
In dimensionless variables, the kinetic energy becomes
\begin{eqnarray}
{\cal W}_{\kappa}(q)=\int_0^q q\,\frac{d q}{\sqrt{1+\kappa^2q^2}} \
\ , \label{RIII17}
\end{eqnarray}
and after introducing the $\kappa$-differential can be written in
the form
\begin{eqnarray}
{\cal W}_{\kappa}(q)=\int_0^q \!q\,\,\,d_{\kappa}q  \ \ ,
\label{RIII18}
\end{eqnarray}
obtaining in this way the well known relativistic expression ${\cal
W}_{\kappa}(q)=(\sqrt{1+\kappa^2q^2}-1)/\kappa^2$.

It is remarkable that the replacement of the ordinary integration by
the $\kappa$-integration, in the classical definition
(\ref{RIII15}), is sufficient to recover the relativistic
expression of the kinetic energy.

Eq. (\ref{RIII18}) linking ${\cal W}$ and $q$ through the
$\kappa$-integral, can be written also in the following differential
form
\begin{eqnarray}
\frac{d }{d_{\kappa} q}\,{\cal W}_{\kappa}(q)= q \ \ ,
\label{RIII19}
\end{eqnarray}
involving the $\kappa$-derivative
\begin{eqnarray}
\frac{d }{d_{\kappa} q}=\sqrt{1+\kappa^2q^2}\,\frac{d}{dq} \ \ .
\label{RIII20}
\end{eqnarray}
Eq. (\ref{RIII19}) after integration with the condition ${\cal
W}_{\kappa}(0)=0$, yields the relativistic expression of the kinetic
energy. In the $\kappa \rightarrow 0$ limit the differential
equation (\ref{RIII19}), reduces to the classical one
\begin{eqnarray}
\frac{d}{d q} \,{\cal W}_{0}(q)= q \ \ . \label{RIII21}
\end{eqnarray}

\subsection{Momentum generalized sum and $\kappa$-differential calculus}

Let us consider in the inertial one-dimension spatial frame $\Sigma$, two
identical, non interacting, free particles $A$ and $B$ of rest mass
$m$ which move toward right. We suppose that the dimensionless
momenta for the two particle are given by $q_{\scriptscriptstyle
A}=q+dq$, $q_{\scriptscriptstyle B}=q$ respectively.

In the rest frame ${\Sigma'}={\Sigma}_{\scriptscriptstyle B}$, of
the particle $B$ the particle momenta are given by
$q'_{\scriptscriptstyle B}=0$ and $q'_{\scriptscriptstyle A}=(q+dq)
\stackrel{\kappa}{\oplus}(-q)$ respectively.

The infinitesimal difference of the particle momenta, in the frame
$\Sigma$ is $dq$, while in the frame $\Sigma'$ is given by
\begin{eqnarray}
d _{\kappa}q=\!\!\!\!&&(q+dq) \stackrel{\kappa}{\oplus}(-q) \\
=\!\!\!\!&&(q+dq) \stackrel{\kappa}{\ominus}q \ \ , \label{MIII1}
\end{eqnarray}
and results to be
\begin{eqnarray}
 d _{\kappa}q= \frac{d\,q}{\displaystyle{\sqrt{1+\kappa^2\,q^2} }} \ .
\label{MIII2}
\end{eqnarray}

In order to better understand the origin  of the expression of the
${\kappa}$-differential, we recall that the variable $q$ is a
dimensionless momentum. Then the quantity
$\gamma(q)=\sqrt{1+\kappa^2\,q^2}$ is the Lorentz factor in the
momentum representation, so that the $\kappa$-differential can be
written as
\begin{eqnarray}
 d _{\kappa}q= \frac{dq}{\gamma(q)} \ .
\label{MIII2a}
\end{eqnarray}

In other words if $dq$ is the infinitesimal difference of two
particle momenta in the inertial frame $\Sigma$, this difference if
observed in rest frame $\Sigma'$ of one of the two particles,
becomes $d _{\kappa}q$ and results to be contracted by the Lorentz
factor.

The $\kappa$-derivative of the function $f(q)$, is defined through
\begin{equation}
\frac{d\,f(q)}{d_{\kappa} \,q}=\lim_{z\rightarrow q}\frac{f
(z)-f(q)}{\displaystyle{ z \stackrel{\kappa}{\ominus}q}}=\frac{f
(q+dq)-f(q)}{\displaystyle{ (q+dq) \stackrel{\kappa}{\ominus}q}}\ \
. \label{MIII4}
\end{equation}

We observe that $df(q)/d_{\kappa}q$, which reduces to $df(q)/dq$ as
the deformation parameter ${\kappa}\rightarrow 0$, can be written in
the form
\begin{eqnarray}
\frac{d \, f(q)}{d _{\kappa}\,q}=\sqrt{1+\kappa^2\,q^2}\,\,\,\frac{d
\, f(q)}{d \, q} \ . \label{MIII5}
\end{eqnarray}
From the latter equation follows that the generalized derivative
obeys the Leibniz's rules of the ordinary derivative. After
introducing the  Lorentz factor $\gamma(q)$, the derivative operator
can be written also in the form:
\begin{eqnarray}
\frac{d }{d _{\kappa}\,q}=\gamma(q)\,\frac{d }{d \, q} \ .
\label{MIII5a}
\end{eqnarray}

The $\kappa$-integral is defined through
\begin{eqnarray}
\int d_{\kappa}q \,\, f(q)= \int \frac{d \, q}{\gamma(q)}\,\,f(q) \
, \label{MIII6}
\end{eqnarray}
and obeys the same rules of the ordinary integral, which recovers
when  ${\kappa}\rightarrow 0$.

\sect{The $\kappa$-exponential function}

\subsection{Definition}

In the present section, the independent variable is a dimensionless
momentum and is indicated by $x$, $y$ or $z$. We recall that the
ordinary exponential $f(x)=\exp(x)$ emerges as solution both of the
functional equation $f(x+y)=f(x)f(y)$ and of the differential
equation $(d/dx)f(x)=f(x)$.

The question to determine the solutions of the generalized equations
\begin{eqnarray}
f(x\stackrel{\kappa}{\oplus}y)=f(x)f(y) \ \ , \label{MIV1}
\end{eqnarray}
\begin{equation}
\frac{d\,f(x)}{d_{\kappa}x}=f(x) \ \ , \label{MIV2}
\end{equation}
reducing in the $\kappa \rightarrow 0$ limit, to the ordinary
exponential, naturally arises. The latter two equations admit the
same solution, which represents an one-parameter generalization of
the ordinary exponential.

{\it Solution of Eq. (\ref{MIV1}):} We write this equation
explicitly
\begin{eqnarray}
f\left(x\sqrt{1+\kappa^2y^2}+y\sqrt{1+\kappa^2x^2}\,\right)=f(x)f(y)
\ \ , \label{MIV3}
\end{eqnarray}
and after performing the change of variables $f(x)=\exp(g(\kappa
x))$, $z_1=\kappa x$, $z_2=\kappa y$ it transforms into the
following equation
\begin{eqnarray}
g\left(z_1\sqrt{1+z_2^2}+z_2\sqrt{1+z_1^2}\,\right)=g(z_1)+g(z_2) \
\ , \label{MIV4}
\end{eqnarray}
which admits the solution $g(x)=A\, {\rm arcsinh} x$. Then, it
results that $f(x)=\exp(A\, {\rm arcsinh} \,\kappa x)$. The
arbitrary constant $A$ can be fixed through the condition
$\lim_{\kappa \rightarrow 0} f(x)=\exp(x)$, obtaining $A=1/\kappa$.
Therefore $f(x)$ assumes the form  $f(x)=\exp_{\kappa}\!\left(x
\right)$ being
\begin{eqnarray}
\exp_{\kappa}(x)= \exp\left( \frac{1}{\kappa} \,{\rm arcsinh}\,
\kappa x \right) \ \ . \label{MIV5}
\end{eqnarray}

{\it Solution of Eq. (\ref{MIV2}):} After performing the change of
variable $\rho={\kappa}^{-1} \,{\rm arcsinh}\, \kappa x $, it
obtains $d_{\kappa}x=d\rho$ and Eq. (\ref{MIV6}) assumes the
form
\begin{equation}
\frac{d\,f}{d\rho}=f \ \ . \label{MIV7}
\end{equation}
From the solution of the latter equation with the condition $f(0)=1$,
follows immediately that $f(x)=\exp_{\kappa}(x)$ with
\begin{eqnarray}
\exp_{\kappa}(x)=\exp \,\left (\frac{1}{\kappa}\, {\rm arcsinh}
\,\kappa x \right) \ \ . \label{MIV8}
\end{eqnarray}
By taking into account that ${\rm arcsinh} \,x =\ln
(\sqrt{1+x^2}+x)$ we can write $\exp_{\kappa}(x)$ in the form
\begin{eqnarray}
\exp_{\kappa}(x)= \left(\sqrt{1+\kappa^{\,2} x^{\,2}}+\kappa
x\right)^{1/\kappa}\ \  , \label{MIV9}
\end{eqnarray}
which will be used in the following.  We remark that
$\exp_{\kappa}(x)$ given by Eq. (\ref{MIV9}), is solution both of
the Eqs. (\ref{MIV1}) and (\ref{MIV2}) and therefore represents a
generalization of the ordinary exponential.

In particular according to Eq. (\ref{MIV2}) the $\kappa$-exponential
is defined as eigenfunction of the $\kappa$-derivative i.e.
\begin{equation}
\frac{d}{d_{\kappa}x}\,\,\exp_{\kappa}(x)=\exp_{\kappa}(x) \ \ .
\label{MIV6} \nonumber
\end{equation}

\subsection{Basic Properties}

From the definition (\ref{MIV9}) of $\exp_{\kappa}(x)$, follows that
\begin{eqnarray}
&&\exp_{\,0}(x)\equiv \lim_{\kappa \rightarrow
0}\exp_{\,\kappa}(x)=\exp (x) \ \ , \label{MIV10}
\\
&&\exp_{- \kappa}(x)=\exp_{\kappa}(x) \ \ .  \label{MIV11}
\end{eqnarray}
Like the ordinary exponential, $\exp_{\kappa}(x)$ has the properties
\begin{eqnarray}
&&\exp_{\kappa}(x) \in C^{\infty}({\bf R}),
\label{MIV12}  \\
&&\frac{d}{d\,x}\, \exp_{\kappa}(x)>0,
\label{MIV13}  \\
&&\exp_{\kappa}(-\infty)=0^+,
\label{MIV14}  \\
&&\exp_{\kappa}(0)=1,
\label{MIV15}  \\
&&\exp_{\kappa}(+\infty)=+\infty,
\label{MIV16}  \\
&&\exp_{\kappa}(x)\exp_{\kappa}(-x)= 1 \ \ . \label{MIV17}
\end{eqnarray}

The property (\ref{MIV17}) emerges as particular case of the more
general one
\begin{eqnarray}
\exp_{\kappa}(x)\exp_{\kappa}(y)=\exp_{\kappa}(x\stackrel{\kappa}{\oplus}y)\
\ . \label{MIV18}
\end{eqnarray}

Furthermore $\exp_{\kappa}(x)$ has the property
\begin{eqnarray}
&&\big[\exp_{\kappa}(x)\big ]^{r} =\exp_{\kappa/r}(r x) \ \ ,
 \label{MIV19}
\end{eqnarray}
with $r\in {\bf R}$, which in the limit $\kappa \rightarrow 0$
reproduces one well known property of the ordinary exponential.

We remark the following convexity property
\begin{eqnarray}
\frac{d^2}{d\,x^2}\, \exp_{\kappa}(x)>0 \ \ ; \ \ x \in {\bf R} \ \
, \label{MIV20}
\end{eqnarray}
holding when $\kappa^2<1$.

Undoubtedly one of the more interesting properties of
$\exp_{\kappa}(x)$, is its power law asymptotic behavior
\begin{eqnarray}
\exp_{\kappa}(x) {\atop\stackrel{\textstyle\sim}{\scriptstyle
x\rightarrow \pm\infty}}\big|\,2\kappa x\big|^{\pm1/|\kappa|} \ \ .
\label{MIV21}
\end{eqnarray}

It is remarkable that the first three terms in the Taylor expansion
of $\exp_{\kappa}(x)$ are the same as those of the ordinary exponential
\begin{equation}
\exp_{\kappa}(x) = 1+ x + \frac{x^2}{2} +
(1-\kappa^2)\,\frac{x^3}{3!}+...  \ \ . \label{MIV52}
\end{equation}
This latter result is a particular case of a more general property
of the relativistic dynamics. Indeed, the Taylor expansion up to the
second order, of any relativistic formula, coincides with the
corresponding classical formula.

\sect{The $\kappa$-logarithm function}

\subsection{Definition}

The function $\ln_{\kappa}(x)$ is defined as the inverse function of
$\exp_{\kappa}(x)$, namely
\begin{eqnarray}
\ln_{\kappa}(\exp_{\kappa}x)=\exp_{\kappa}(\ln_{\kappa}x)=x \ ,
\label{MV1}
\end{eqnarray}
and is given by
\begin{eqnarray}
\ln_{\kappa}(x) = \frac{1}{\kappa}\,\sinh \, (\kappa \ln x)\ \ ,
\label{MV3}
\end{eqnarray}
or more properly
\begin{eqnarray}
\ln_{\kappa}(x)= \frac{x^{\kappa}-x^{-\kappa}}{2\kappa} \ \ .
\label{MV4}
\end{eqnarray}

\subsection{Basic properties}

It results that
\begin{eqnarray}
&&\ln_{0}(x)\equiv \lim_{\kappa \rightarrow 0}\ln_{\kappa}(x)=\ln
(x) \ \ , \label{MV4}
\\
&&\ln_{- \kappa}(x)=\ln_{\kappa}(x) \ \ . \label{MV5}
\end{eqnarray}

The function $\ln_{\kappa}(x)$, just as the ordinary logarithm, has
the properties
\begin{eqnarray}
&&\ln_{\kappa}(x) \in C^{\infty}({\bf R}^+),
\label{MV6} \\
&&\frac{d}{d\,x}\, \ln_{\kappa}(x)>0,
\label{MV7}  \\
&&\ln_{\kappa}(0^+)=-\infty,
\label{MV8}  \\
&&\ln_{\kappa}(1)=0,
\label{MV9}  \\
&&\ln_{\kappa}(+\infty)=+\infty,
\label{MV10}  \\
&&\ln_{\kappa}(1/x)=-\ln_{\kappa}(x) \ \ . \label{MV11}
\end{eqnarray}

Furthermore $\ln_{\kappa}(x)$ has the two properties
\begin{eqnarray}
&&\ln_{\kappa}(x^{r}) =r \ln_{r \kappa}(x) \ \ ,  \label{MV12}
\\
&&\ln_{\kappa}(x \,y) =\ln_{\kappa}(x)
\oplus\!\!\!\!\!^{^{\scriptstyle \kappa}}\,\,\ln_{\kappa}(y)\ \ ,
 \label{MV13}
\end{eqnarray}
with $r\in {\bf R}$. Note that the property (\ref{MV11}) follows as
particular case of the property (\ref{MV12}).

We remark the following concavity properties
\begin{eqnarray}
&&\frac{d^2}{d\,x^2}\, \ln_{\kappa}(x)<0
 \ \ , \label{MV14}   \\
&&\frac{d^2}{d\,x^2}\,x\, \ln_{\kappa}(x)<0  \ \ . \label{MV15}
\end{eqnarray}

A very interesting property of this function is its power law
asymptotic behavior
\begin{eqnarray}
&&\ln_{\kappa}(x) {\atop\stackrel{\textstyle\sim}{\scriptstyle
x\rightarrow
0^+}}-\frac{1}{2\,|\kappa|}\,x^{-|\kappa|} \ \ , \label{MV16}  \\
&&\ln_{\kappa}(x) {\atop\stackrel{\textstyle\sim}{\scriptstyle
x\rightarrow +\infty}}\,\,\frac{1}{2\,|\kappa|}\,x^{|\kappa|} \ \ .
\label{MV17}
\end{eqnarray}

After recalling the integral representation of the ordinary
logarithm
\begin{equation}
\ln(x) = \frac{1}{2}\,\int_{1/x}^{x}\frac{1}{t} \,dt \ \ , \nonumber
\label{MV18}
\end{equation}
one can verify that the latter relationship can be generalized
easily in order to obtain $\ln_{\kappa}(x)$,  by replacing the
integrand function $y_0(t)=t^{-1}$ by the new function
$y_{\kappa}(t)=t^{-1-\kappa}$, namely
\begin{equation}
\ln_{\kappa}(x) =
\frac{1}{2}\,\int_{1/x}^{x}\,\frac{1}{t^{1+\kappa}} \,dt \ \ .
\nonumber  \label{MV19}
\end{equation}

The first terms of the Taylor expansion related to
$\kappa$-logarithm, are
\begin{equation}
\ln_{\kappa}(1+x) = x - \frac{x^2}{2} +
\left(1+\frac{\kappa^2}{2}\right)\frac{x^3}{3} - ... \ \ . \nonumber
\label{MV20}
\end{equation}

\subsection{The $\ln_{\kappa}(x)$ as solution of functional equations}

{\it First functional equation:} The logarithm $f(x)=\ln(x)$ is the
only existing function, except for a multiplicative constant, which
results to be solution of the function equation
$f(x_1x_2)=f(x_1)+f(x_2)$. Let us consider now the generalization of
this equation,  obtained by substituting the ordinary sum by the
momentum generalized sum
\begin{eqnarray}
f(x_1x_2)= f(x_1)\stackrel{\kappa}{\oplus} f(x_2) \ \ . \label{MV21}
\end{eqnarray}
We proceed by solving this equation, which assumes the explicit form
\begin{eqnarray}
f(x_1x_2)&=& f(x_1)\,\sqrt{1+\kappa^2\,f(x_2)\,^2} \nonumber \\
&+& f(x_2)\,\sqrt{1+\kappa^2\,f(x_1)\,^2} \ \ \ . \label{MV22}
\end{eqnarray}
After performing the substitution $f(x)=\kappa^{-1} \sinh \kappa
g(x)$ we obtain that the auxiliary function $g(x)$ obeys the
equation $g(x_1x_2)=g(x_1)+g(x_2)$, and then is given by $g(x)=A\ln
x$. The unknown function becomes $f(x)=\kappa^{-1}\sinh ( \kappa \ln
x)$ where we have set $A=1$ in order to recover, in the limit
$\kappa\rightarrow 0$, the classical solution $f(x)=\ln(x)$. Then we
can conclude that the solution of Eq. ($\ref{MV21}$) is given by
\begin{eqnarray}
f(x)=\ln_{\kappa}(x) \ \ . \label{MV23}
\end{eqnarray}

{\it Second functional equation:} The following first order
differential-functional equation emerges in statistical mechanics
within the context of the maximum entropy principle
\begin{eqnarray}
&&\frac{d}{dx}\,[\,x\,\,f\,(x)\,]=\lambda \,f\left(x/\alpha\right) \
\ , \label{MV24}
\\&&f(1)=0 \ \ , \label{MV25}
\\ &&f'(1)=1 \ \ , \label{MV26}
\\ &&f\left(1/x\right)=-f\left(x\right), \label{MV27}
\end{eqnarray}
$\alpha$ and $\lambda$ being two arbitrary constants. The latter
problem admits two solutions. The first is given by $f(x)=\ln (x)$
and $\alpha=1/e$, $\lambda=1$. The second solution is obtained after
tedious but straightforward calculations \cite{PRE02}, and is given
by
\begin{eqnarray}
f(x)= \ln_{\kappa}(x) \ , \ \ \ \label{MV28}
\end{eqnarray}
and
\begin{eqnarray}
&&\alpha=\left(\frac{1-\kappa}{1+\kappa}\right)^{1/2\kappa} \ ,
\label{MV29} \\
&&\lambda=\sqrt{1-\kappa^2} \ . \label{MV30}
\end{eqnarray}

\subsection{The Entropy}

A physically meaningful link between the functions $\ln_{\kappa}(x)$
and $\exp_{\kappa}(x)$ is given by the following variational
principle.

Let be $h(q)$ an arbitrary real function and $f(q)$ a real positive
function of the variable $q\in A$. The solution of the variational
equation
\begin{equation}
\frac{\delta}{\delta f(q)}\left[-\int_Adq \,\,f(q)\ln_{\kappa}f(q)
+\int_Adq \,\,f(q)\,h(q) \right]=0 \ , \label{MV31}
\end{equation}
is unique and is given by
\begin{eqnarray}
f(q)=\alpha \, \exp_{\kappa}\!\big(h(q)/\lambda\big) \ ,
\label{MV32}
\end{eqnarray}
$\alpha$ and $\lambda$ being the constants defined by Eqs.
(\ref{MV29}) and (\ref{MV30}). The solution  of the variational
equation (\ref{MV31}) is trivial and employs Eq. (\ref{MV24}).

This important result permits us to interpret the functional
\begin{eqnarray}
S_{\kappa}=-\int_Adq \,\,f(q)\ln_{\kappa}f(q) \ , \label{MV33}
\end{eqnarray}
which can be written also in the form
\begin{eqnarray}
S_{\kappa}=\int_Adq
\,\,\,\frac{f(q)^{1-\kappa}-f(q)^{1+\kappa}}{2\kappa} \ ,
\label{MV34}
\end{eqnarray}
as the entropy associated to the function
$\exp_{\kappa}\left(x\right)$. It is remarkable that in the $\kappa
\rightarrow 0$ limit, as $\ln_{\kappa}(x)$ and
$\exp_{\kappa}\left(x\right)$ approach $\ln (x)$ and $\exp(x)$
respectively, the new entropy reduces to the old Boltzmann-Shannon
entropy.

It is shown that the entropy $S_{\kappa}$ has the standard
properties of Boltzmann-Shannon entropy: is thermodynamically
stable, is Lesche stable, obeys the Khinchin axioms of continuity,
maximality, expandability and generalized additivity.

\sect{Other Representations of the $\kappa$-exponential}

\subsection{Rapidity variable}

We define the rapidity as $r=c\,\,{\rm arctanh} \,(v/c)$, so that in
the classical limit it reduces to the particle velocity. The rapidity
in dimensionless form $\rho=r/c\kappa$, is given by
\begin{eqnarray}
\rho(u)=\frac{1}{\kappa}\,{\rm arctanh}(\kappa u) \ \ . \
\label{RVI1}
\end{eqnarray}
Clearly the above expression of the rapidity is written in terms of
the velocity, but it can be written also in terms of other
variables. For instance, after taking into account the standard
formulas of relativistic dynamics e.g. $u=q/\sqrt{1+\kappa^2q^2}$
etc, the rapidity can be expressed in terms of the momentum, of the
total energy, and of the kinetic energy, according to
\begin{eqnarray}
&&\rho(q)=\frac{1}{\kappa}\,{\rm arcsinh} (\kappa q)
\label{RVI2}\\
&&\rho({\cal E})=\frac{1}{\kappa}\,{\rm arccosh}(\kappa^2{\cal E})
\label{RVI3}\\
&&\rho({\cal W})=\frac{1}{\kappa}\,{\rm arccosh}(1+\kappa^2{\cal W})
\ \ , \ \label{RVI4}
\end{eqnarray}
and obviously it results in
\begin{eqnarray}
\rho=\rho(q)=\rho(u)=\rho({\cal E})=\rho({\cal W}) \ \ .
\label{RVI5}
\end{eqnarray}

\subsection{The function $\kappa$-exponential
in the velocity representation}

We recall that the composition laws of the momenta and of the
velocities are two generalized sums. For this reason we will
consider more in detail the rapidity variable in the momentum
representation and in the velocity representation. From Eq.
(\ref{RVI5}) we have
\begin{eqnarray}
\exp(\rho)=\exp(\rho(q))=\exp(\rho(u)) \ \ , \label{RVI6}
\end{eqnarray}
or equivalently
\begin{eqnarray}
\exp(\rho)=\exp_{\kappa}(q)=\exp^{\kappa}(u) \ \ . \label{RVI7}
\end{eqnarray}
The functions $\exp_{\kappa}(q)$ and $\exp^{\kappa}(u)$ are defined
through
\begin{eqnarray}
&&\exp_{\kappa}(q)=\exp \,\left (\frac{1}{\kappa}\, {\rm arcsinh}
\,\kappa q \right) \ \ , \label{RVI8} \\
&&\exp^{\kappa}(u)=\exp \,\left (\frac{1}{\kappa}\, {\rm arctanh}
\,\kappa u \right) \ \ , \label{RVI9}
\end{eqnarray}
or equivalently
\begin{eqnarray}
&&\exp_{\kappa}(q)=\left(\sqrt{1+\kappa^2 q^2}+\kappa q
\right)^{1/\kappa} \ \ , \label{RVI10} \\
&&\exp^{\kappa}(u)=\left(\frac{1+\kappa u}{1-\kappa u}
\right)^{1/2\kappa} \ \ . \label{RVI11}
\end{eqnarray}

The explicit relationships linking $\exp_{\kappa}(q)$, and
$\exp^{\kappa}(u)$ are given by
\begin{eqnarray}
&&\exp^{\kappa}(u)=\exp_{\kappa}\big(u\,\gamma(u)\big) \ \ , \label{RVI12} \\
&&\exp_{\kappa}(q)=\exp^{\kappa}\big(q/\gamma(q) \big) \ \ ,
\label{RVI13}
\end{eqnarray}
$\gamma$ being the Lorentz factor linking momentum and velocity
through $q=\gamma\,u$, which has the following expressions in the
velocity representation and in the momentum representation
\begin{eqnarray}
&&\gamma(u)=\frac{1}{\sqrt{1-\kappa^2u^2}} \ \ , \label{RVI14} \\
&&\gamma(q)=\sqrt{1+\kappa^2q^2} \ \ , \label{RVI15}
\end{eqnarray}
respectively.

We can conclude that the three functions $\exp(\rho)$,
$\exp_{\kappa}(q)$, and $\exp^{\kappa}(u)$ are the same function in
three different representations.

Hereafter we discuss briefly the function $\exp^{\kappa}(u)$ which
has been introduced firstly in the appendix of ref. \cite{PRE05}.
Starting from the velocity generalized sum  we can obtain easily the
$\kappa$-differential  in the velocity representation as follous
\begin{eqnarray}
d^{\kappa}u= \frac{du}{1-\kappa^2u^2}\ \ . \label{RVI16}
\end{eqnarray}
The $\kappa$-derivative and the $\kappa$-integral of the function
$f(u)$ with respect the dimensionless velocity, assume respectively
the forms
\begin{eqnarray}
\frac{d\, f(x)}{d^{\kappa}u}= (1-\kappa^2u^2)\,\frac{d\, f(u)}{du} \
\ , \label{RVI17}
\end{eqnarray}
\begin{eqnarray}
\int d^{\kappa}u\,\,f(u)=\int \frac{du}{1-\kappa^2u^2}\, f(u) \ \ .\
\ \label{RVI18}
\end{eqnarray}

The function $f(u)=\exp^{\kappa}(u)$ can be obtained as solution of
the two following equations
\begin{eqnarray}
f(u_1\stackrel{\kappa}{\oplus}u_2)=f(u_1)f(u_2) \ \ , \label{RVI19}
\end{eqnarray}
\begin{equation}
\frac{d\,f(u)}{d^{\kappa}u}=f(u) \ \ . \label{RVI20}
\end{equation}

It is easy to verify that $\exp^{\kappa}(u)$ is a monotonic,
continuous function and is defined in the interval
$-1/|\kappa|<u<1/|\kappa|$. The inverse function of
$\exp^{\kappa}(u)$ namely $\ln^{\kappa}(x)$ is given by
\begin{eqnarray}
&&\ln^{\kappa}(x)= \frac{1}{\kappa}\,\, \tanh (\kappa \ln x) \ \ ,
\label{RVI21} \\
&&\ln^{\kappa}(x)= \frac{1}{\kappa}\,\,
\frac{x^{\kappa}-x^{-\kappa}}{x^{\kappa}+x^{-\kappa}} \ \ .
\label{RVI22}
\end{eqnarray}

The relationships linking $\ln^{\kappa}(x)$ and $\ln_{\kappa}(x)$
are the following
\begin{eqnarray}
&&\ln^{\kappa}(x)= \frac{\ln_{\kappa}(x)}
{\sqrt{1+\kappa^2\left[\,\ln_{\kappa}(x)\right]^2}} \ \ ,
\label{RVI23} \\
&&\ln_{\kappa}(x)= \frac{\ln^{\kappa}(x)}
{\sqrt{1-\kappa^2\left[\,\ln^{\kappa}(x)\right]^2}} \ \ .
\label{RVI24}
\end{eqnarray}

The properties of the functions $\exp^{\kappa}(x)$ and
$\ln^{\kappa}(x)$ follows easily from the ones of the functions
$\exp_{\kappa}(x)$ and $\ln_{\kappa}(x)$. For instance, we obtain
\begin{eqnarray}
&&\exp^{\kappa}(x)\exp^{\kappa}(-x)= 1 \ \ ,
\label{RVI25} \\
&&\ln^{\kappa}(1/x)= - \ln^{\kappa}(x) \ \ , \label{RVI26}
\end{eqnarray}
and so on.

In ref. \cite{PRE02}  a general procedure to deform
the ordinary mathematics starting from an arbitrary generalized sum has been proposed.
This procedure have been adopted in ref. \cite{PhA01} to construct
the deformed mathematics related to the momentum generalized sum and
then, based on the function $\exp_{\kappa}(q)$.

Clearly starting from the velocity generalized sum we can construct
the related deformed mathematics based on the function
$\exp^{\kappa}(u)$.

For instance we can introduce the deformed hyperbolic functions
$\sinh^{\kappa}(u)$ and $\cosh^{\kappa}(u)$ according to
\begin{eqnarray}
&&\sinh^{\kappa}(u)=\frac{\exp^{\kappa}(u)-\exp^{\kappa}(-u)}{2} \ \
,
\label{RVI27} \\
&&\cosh^{\kappa}(u)=\frac{\exp^{\kappa}(u)+\exp^{\kappa}(-u)}{2} \ \
. \label{RVI28}
\end{eqnarray}
These two functions defines a $\kappa$-deformed hyperbolic
trigonometry which is isomorphic with respect the ordinary
hyperbolic trigonometry.

The deformed cyclic functions $\sin^{\kappa}(u)$ and
$\cos^{\kappa}(u)$ given by
\begin{eqnarray}
&&\sin^{\kappa}(u)=\frac{\exp^{\kappa}(iu)-\exp^{\kappa}(-iu)}{2i} \
\ ,
\label{RVI29} \\
&&\cos^{\kappa}(u)=\frac{\exp^{\kappa}(iu)+\exp^{\kappa}(-iu)}{2} \
\ , \label{RVI30}
\end{eqnarray}
defines the $\kappa$-deformed cyclic trigonometry which is
isomorphic with respect the ordinary cyclic trigonometry.

The three mathematical structures based on the ordinary sum, on the
momentum generalized sum, and on the velocity generalized sum,
result to be mutually isomorphic.

Regarding the possibility to consider further representations of the
exponential function in special relativity we recall that the
composition laws of the total energy $\cal E$, and of the kinetic
energy $\cal W$, are not generalized sums and for these reason we
can't use the variables $\cal E$ and $\cal W$ to introduce new
representation of the exponential function. Let us consider for
instance the function
\begin{eqnarray}
f({\cal W})=\exp\big( \,\rho({\cal W})\big)=\exp_{\kappa}\big(
\,q({\cal W})\big) \ \ . \label{RV31}
\end{eqnarray}
Clearly this function can't be considered as a representation of the
exponential function. Indeed it results in
\begin{eqnarray}
f({\cal W})\,f(-{\cal W})\neq 1 \ \ . \label{RVI32}
\end{eqnarray}

In general, a deformation of the ordinary exponential i.e.
$\exp\big(h(x)\big)$ is not an its representation. In order to have
a new representation of the exponential (the mathematics underlying
the new exponential, must be isomorphic with respect the ordinary
mathematics, underlying the ordinary exponential) the function
$h(x)$ must have specific properties \cite{PRE02}.

It is important to note that in the classical limit, the composition
law for the relativistic kinetic energies, given by Eq.
(\ref{RII16}), reduces to the following expression of Galileian
relativity
\begin{eqnarray}
{\cal W}_1\stackrel{0}{\oplus}{\cal W}_2 ={\cal W}_1+{\cal
W}_2+2\sqrt{{\cal W}_1{\cal W}_2} \ \ . \label{RVI33}
\end{eqnarray}
Then, a non trivial composition law, which is not a generalized sum,
appears also in classical physics when we change the particle
observation inertial frame. The latter composition law never has
been used in classical physics to introduce new representations of
the exponential function. We stress that in classical physics, the
ordinary exponential can be viewed as a function emerging in the
momentum or equivalently in the velocity representation. Indeed,
according to the Galileian relativity the composition laws for the
classical momenta and velocities are the ordinary sum and this
ordinary sum generates the ordinary exponential and the ordinary
mathematics.

In conclusion in special relativity emerge only two non trivial
representations of the exponential function namely the deformed
exponentials $\exp_{\kappa}(q)$ and $\exp^{\kappa}(u)$ defined by
Eqs. (\ref{RVI10}) and (\ref{RVI11}) respectivelly. There are
several reasons to select the function $\exp_{\kappa}(q)$ as the
more proper representation of the exponential function in order to
construct the relativistic statistical mechanics.

A first reason is related to the fact that the momentum is the more
proper variable to formulate the one-particle relativistic dynamics.
Indeed the dynamical Lorentz transformations involve the momentum
and no the velocity. Also the relativistic Newton equation in terms
of the momentum, assumes a very simple form, similar to the
classical Newton equation. On the contrary the relativistic Newton
equation, if expressed in terms of velocity and acceleration,
results to be a non-linear equation very different with respect the classical
Newton equation.

A second reason is related to the fact that also in the formulation
of the many body relativistic theory the momentum results to be the
more proper variable. For instance in relativistic kinetics but also
in relativistic field theories, the Lorentz invariant integration
involves the momentum and no the velocity.

The third and more important reason is related to the maximum
entropy principle, the cornerstone of statistical mechanics. Indeed,
the couple of functions $\exp_{\kappa}(q)$ and $\ln_{\kappa}(q)$ are
linked through the variational principle described in the subsection
D of the section V. On the contrary the functions $\exp^{\kappa}(q)$
and $\ln^{\kappa}(q)$ can't be connected by the same variational
principle.

\sect{Relativistic kinetics}

\subsection{The $\kappa$-Product and ${\kappa}$-Sum of functions}

Let us consider the set of the non negative real functions ${\cal
D}=\{f,h,w,...\}$.

{\it Proposition 1:} The composition law
$\otimes\mbox{\raisebox{-2mm}{\hspace{-2.5mm}$\scriptstyle
\kappa$}}\hspace{1mm}$ defined through
\begin{eqnarray}
\ln_{\kappa}(f\otimes\mbox{\raisebox{-2mm}{\hspace{-3.3mm}$\scriptstyle
\kappa$}} \hspace{2mm}h)= \ln_{\kappa}f+\ln_{\kappa}h \ \ ,
\label{VII1}
\end{eqnarray}
which reduces to the ordinary product as $\kappa\rightarrow 0$,
namely $f\otimes\mbox{\raisebox{-2.3mm}{\hspace{-2.7mm}$\scriptstyle
0$}} \hspace{2mm}h= f\cdot h$, is a generalized product and the
algebraic structure $({\cal D}-\{0\},
\otimes\mbox{\raisebox{-2mm}{\hspace{-2.5mm}$\scriptstyle
\kappa$}}\hspace{1mm})$ forms an abelian group.

{\it Proof:} Indeed this $\kappa$-product has the following
properties
\\1) associative law:
$(f\otimes\mbox{\raisebox{-2mm}{\hspace{-3.3mm}$\scriptstyle
\kappa$}}
\hspace{2mm}h)\otimes\mbox{\raisebox{-2mm}{\hspace{-3.3mm}$\scriptstyle
\kappa$}}
\hspace{2mm}w=f\otimes\mbox{\raisebox{-2mm}{\hspace{-3.3mm}$\scriptstyle
\kappa$}} \hspace{2mm}
(h\otimes\mbox{\raisebox{-2mm}{\hspace{-3.3mm}$\scriptstyle
\kappa$}} \hspace{2mm}w)$;
\\ 2) neutral element:
$f\otimes\mbox{\raisebox{-2mm}{\hspace{-3.3mm}$\scriptstyle
\kappa$}}
\hspace{2mm}1=1\otimes\mbox{\raisebox{-2mm}{\hspace{-3.3mm}$\scriptstyle
\kappa$}} \hspace{2mm}f=f$;
\\ 3) inverse element:
$f\otimes\mbox{\raisebox{-2mm}{\hspace{-3.3mm}$\scriptstyle
\kappa$}} \hspace{2mm}(1/f)=
(1/f)\otimes\mbox{\raisebox{-2mm}{\hspace{-3.3mm}$\scriptstyle
\kappa$}} \hspace{2mm}f=1$;
\\ 4) commutative law: $f
\otimes\mbox{\raisebox{-2mm}{\hspace{-3.3mm}$\scriptstyle \kappa$}}
\hspace{2mm}h=h\otimes\mbox{\raisebox{-2mm}{\hspace{-3.3mm}$\scriptstyle
\kappa$}} \hspace{2mm}f$.

Of course the $\kappa$-division
$\oslash\mbox{\raisebox{-2mm}{\hspace{-2.2mm}$\scriptstyle
\kappa$}}\hspace{.5mm}$ can be defined as follows
$f\oslash\mbox{\raisebox{-2mm}{\hspace{-3.3mm}$\scriptstyle
\kappa$}}\hspace{2mm}h=f
\otimes\mbox{\raisebox{-2mm}{\hspace{-3.3mm}$\scriptstyle \kappa$}}
\hspace{1mm}\,(1/h)$.

The deformed $\kappa$-power $f^{\otimes r}$ is defined through
\begin{equation}
\ln_{\kappa}\big(f^{\otimes r}\,\big)=r\,\ln_{\kappa}f \ \ ,
\label{VII2}
\end{equation}
and generalizes the ordinary power $f^r$. In particular, when $r$ is
integer one has $f^{\otimes r}=
f\otimes\mbox{\raisebox{-2mm}{\hspace{-3.3mm}$\scriptstyle \kappa$}}
\hspace{2mm}f ...
\otimes\mbox{\raisebox{-2mm}{\hspace{-3.3mm}$\scriptstyle \kappa$}}
\hspace{2mm}f$, (r times).

The $\kappa$-product allows us to write the following property of
the $\kappa$-exponential
\begin{eqnarray} \exp_{\kappa}(x)\otimes\mbox{\raisebox{-2mm}{\hspace{-3.3mm}$\scriptstyle
\kappa$}} \hspace{2mm} \!\exp_{\kappa}(y)=\exp_{\kappa}(x+y) \ .
\label{VII3}
\end{eqnarray}

{\it Proposition 2:} The algebraic structure $({\cal D},
\otimes\mbox{\raisebox{-2mm}{\hspace{-2.5mm}$\scriptstyle
\kappa$}}\hspace{1mm})$ forms an abelian monoid. \\ {\it Proof:}
Indeed the element $0$ does not admit an inverse element.
\\
Furthermore, just as in the case of the ordinary product, it results
$f\otimes\mbox{\raisebox{-2mm}{\hspace{-3.3mm}$\scriptstyle
\kappa$}}
\hspace{2mm}0=0\otimes\mbox{\raisebox{-2mm}{\hspace{-3.3mm}$\scriptstyle
\kappa$}} \hspace{2mm}f= 0$.

{\it Proposition 3:} The composition law
$\oplus\mbox{\raisebox{-2mm}{\hspace{-2.5mm}$\scriptstyle \kappa$}}
\hspace{1mm}$ defined through
\begin{equation}
\exp\Big(\ln_{\kappa}\big(f\oplus\mbox{\raisebox{-2mm}{\hspace{-3.3mm}$\scriptstyle
\kappa$}} \hspace{2mm}h\big)\Big) =
\exp\left(\,\ln_{\kappa}f\right)+ \exp\left(\,\ln_{\kappa}h\right) \
, \label{VII4}
\end{equation}
which reduces to the ordinary sum as the deformation parameter
approaches to zero, namely $f
\oplus\mbox{\raisebox{-2.4mm}{\hspace{-3.3mm}$\scriptstyle0$}}
\hspace{2mm}h=f+h$, is a generalized sum and the algebraic structure
$({\cal D},\otimes\mbox{\raisebox{-2mm}{\hspace{-2.5mm}$\scriptstyle
\kappa$}}\hspace{1mm})$ forms an abelian monoid.

{\it Proof:} Indeed this $\kappa$-sum has the following properties
\\ 1) associative
law: $(f\oplus\mbox{\raisebox{-2mm}{\hspace{-3.3mm}$\scriptstyle
\kappa$}} \hspace{2mm}h)
\oplus\mbox{\raisebox{-2mm}{\hspace{-3.3mm}$\scriptstyle \kappa$}}
\hspace{2mm}w=f\oplus\mbox{\raisebox{-2mm}{\hspace{-3mm}$\scriptstyle
\kappa$}} \hspace{2mm}(h
\oplus\mbox{\raisebox{-2mm}{\hspace{-3.3mm}$\scriptstyle \kappa$}}
\hspace{2mm}w)$;\\ 2) neutral element:
$f\oplus\mbox{\raisebox{-2mm}{\hspace{-3.3mm}$\scriptstyle \kappa$}}
\hspace{2mm}0=0
\oplus\mbox{\raisebox{-2mm}{\hspace{-3.3mm}$\scriptstyle \kappa$}}
\hspace{2mm}f=f$;\\ 3) commutative law:
$f\oplus\mbox{\raisebox{-2mm}{\hspace{-3.3mm}$\scriptstyle \kappa$}}
\hspace{2mm}h=h\oplus\mbox{\raisebox{-2mm}{\hspace{-3.3mm}$\scriptstyle
\kappa$}} \hspace{2mm}f$.

{\it Proposition 4:} The product
$\otimes\mbox{\raisebox{-2mm}{\hspace{-2.5mm}$\scriptstyle \kappa$}}
\hspace{1mm}$ and sum
$\oplus\mbox{\raisebox{-2mm}{\hspace{-2.4mm}$\scriptstyle \kappa$}}
\hspace{1mm}$ are distributive operations
\begin{eqnarray}
w\otimes\mbox{\raisebox{-2mm}{\hspace{-2.3mm}$\scriptstyle \kappa$}}
\hspace{2mm}(f\oplus\mbox{\raisebox{-2mm}{\hspace{-3.3mm}$\scriptstyle
\kappa$}} \hspace{2mm}h)=
(w\otimes\mbox{\raisebox{-2mm}{\hspace{-3.3mm}$\scriptstyle
\kappa$}} \hspace{2mm}f)
\oplus\mbox{\raisebox{-2mm}{\hspace{-3.2mm}$\scriptstyle \kappa$}}
\hspace{2mm}(w\otimes\mbox{\raisebox{-2mm}{\hspace{-3.5mm}$\scriptstyle
\kappa$}} \hspace{2mm}h) \ \ . \label{VII5}
\end{eqnarray}

\subsection{Evolution Equation}

By using the standard notations of the relativistic theory we denote
with $x=x^{\nu}=(ct,\mbox{\boldmath $x$})$ the four-vector position
and with $p=p^{\nu}=(p^{0},\mbox{\boldmath $p$})$ the four-vector
momentum, being $p^{0}=\sqrt{\mbox{\boldmath $p$}^2 + m^2c^2}$ and
employ the metric $g^{\mu\nu}=diag\,(1,-1,-1,-1)$ \cite{DEGROOT}.

Let us consider the following relativistic kinetic equation
\begin{eqnarray}
p^{\,\nu}\partial_{\nu}f-m F^{\nu}\frac{\partial f}{\partial
p^{\,\nu}}= \int
\frac{d^3p'}{{p'}^{0}}\frac{d^3p_1}{p_1^{\,0}}\frac{d^3p'_1}{{p'}_{\!\!1}^{0}}
\,\,G \,\,&& \nonumber \\
\times \left[\,C(f',f'_1)- C(f,f_1) \,\right],&& \label{VII6}
\end{eqnarray}
where the distribution $f=f(x,p)$ is a function of the four-vectors
position and momentum, $G$ is the transition rate which depends only
on the nature of the two body particle interaction and $C(f,f_1)$ is
the two particle correlation function with the same four-vector
position $x$, and four-momenta $p$ and $p_1$ respectively.

We note that the left hand side of Eq. (\ref{VII6}) is the same as in
the standard relativistic Boltzmann equation. In the particular case
where the two particle correlation function is assumed to have the
same expression like in the classical Boltzmann equation i.e.
$C(f,f_1)=f\,f_1$ (Stosszahlansatz), the above equation reduces to
the ordinary relativistic Boltzmann equation \cite{DEGROOT},
admitting as stationary distribution an exponential distribution,
known as relativistic Maxwell-Boltzmann distribution or as Juttner
distribution.

Clearly in the case where $C(f,f_1)\neq f\,f_1$, Eq. (\ref{VII6})
describes a new relativistic kinetics, radically different from the
standard one. In the following we pose

\begin{eqnarray}
C(f,f_1)=
(f/\alpha)\,\,\otimes\mbox{\raisebox{-2mm}{\hspace{-3.3mm}$\scriptstyle
\kappa$}} \hspace{1mm}\,\,(f_1/\alpha)  \ \ . \label{VII7}
\end{eqnarray}
The origin and value of the constant $\alpha$ will be discussed in
the following.

\subsection{Stationary distribution}

We consider now the steady states of Eq. (\ref{VII6}) for which the
collision integral becomes equal to zero. Then we have
\begin{eqnarray}
(f/\alpha)\otimes\mbox{\raisebox{-2.3mm}{\hspace{-3.3mm}$\scriptstyle
\kappa$}}\hspace{1mm}(f_1/\alpha)=
(f'/\alpha)\otimes\mbox{\raisebox{-2mm}{\hspace{-4.8mm}
$\scriptstyle\kappa$}} \hspace{1mm}\,(f'_1/\alpha) \ \ ,
\label{VII8}
\end{eqnarray}
and after taking into account the definition (\ref{VII1}) of the
$\kappa$-product, we obtain
\begin{equation}
\ln_{\kappa}(f/\alpha)+ \ln_{\kappa}(f_{1}/\alpha)
=\ln_{\kappa}(f^\prime/\alpha) +\ln_{\kappa}(f^\prime_{1}/\alpha) \
. \ \ \label{VII9}
\end{equation}
This last equation represents a conservation law and then we can
conclude that $\ln_{\kappa}(f/\alpha)$ is a summational invariant;
in the most general case it is a linear combination of the
microscopic relativistic invariants, namely a constant and the
four-vector momentum. In the literature  it is shown that in
presence of external electromagnetic fields the more general
microscopic relativistic invariant has a form proportional to
$\left(p^{\nu}+q A^{\nu}\!/c \right)\,U_{\nu}+$ constant, being
$U_{\nu}$  the hydrodynamic four-vector velocity with
$U^{\nu}U_{\nu}=c^2$. Then we can pose
\begin{equation}
\ln_{\kappa}(f/\alpha)=-\frac{\beta}{\lambda}\,\big[\left(p^{\nu}+q
A^{\nu}\!/c \right)\,U_{\nu}-mc^2\,\big] + \frac{\beta\mu}{\lambda}
 \ . \label{VII10}
\end{equation}
Consequently we obtain the following stationary distribution
\begin{equation}
f=\alpha \exp_{\kappa}\bigg(-\beta\,\frac{\left(p^{\nu}+q
A^{\nu}\!/c \right)\,U_{\nu}-mc^2-\mu }{\lambda}\bigg) \ .
\label{VII11}
\end{equation}

At the moment $\alpha$, $\lambda$, $\beta$, and $\mu$ remains
arbitrary constants which will be calculated and/or interpreted by
using the Maximum Entropy Principle imposing that the stationary
distribution of the system (\ref{VII11}) must maximize the entropy
of the system.

\subsection{The maximum Entropy Principle}

We define the four-vector entropy $S^{\nu}=(S^{0},\mbox{\boldmath
$S$})$ as follows
\begin{equation}
S^{\nu}= - \int \frac{d^3p}{p^{0}}\,p^{\nu}\, f \,\ln_{\kappa}\!f \
. \label{VII12}
\end{equation}

The identity $d^3p/p^0=d^4p \,\, 2\,
\theta(p^0)\,\delta(p^{\mu}p_{\mu}-m^2c^2)$ permits us to write
$S^{\nu}$ also in the form
\begin{equation}
S^{\nu}= - \int
d^4p\,\,2\,\theta(p^{0})\,\delta(p^{\mu}p_{\mu}-m^2c^2)\,\,p^{\nu}\,
f \,\ln_{\kappa}\!f \ \ . \label{VII13}
\end{equation}
In the latter expression $d^4p$ is a scalar because the Jacobian of
the Lorentz transformation is equal to unit. Then since $p^{\nu}$
transforms as a four-vector, we can conclude that $S^{\nu}$
transforms as a four-vector.

The quantity $\mbox{\boldmath $S$}$ is the entropy flow while
$S^{0}=S$ is the $\kappa$-entropy given by
\begin{equation}
S= - \int d^3p \, f \,\ln_{\kappa}\!f \ . \label{VII14}
\end{equation}

The maximization of the latter entropy under the constraints
imposing the conservation of the norm of the distribution $f$, and
the a priori knowledge of the value of the more general microscopic
invariant, conducts to the following variational equation

\begin{eqnarray}
&&\frac{\delta}{\delta f}\,\,\Bigg\{- \int d^3p \,\, f
\,\ln_{\kappa}\!f + \beta \mu \int d^3p \,\,\, f \,- \nonumber  \\
&&- \beta \int d^3p \,\,\big[\left(p^{\nu}+q A^{\nu}\!/c
\right)\,U_{\nu} -mc^2\big]\,f\,\Bigg\}=0 \ , \ \ \ \ \ \ \ \ \ \
\label{VIII15}
\end{eqnarray}
$\beta$ and $\mu$ being the Lagrange multipliers.

The solution of the latter variational problem conduct to the
stationary distribution (\ref{VII11}) only thanks to the fact that
the function $\ln_{\kappa}(f)$ has the property
\begin{eqnarray}
\frac{d}{df}\,[\,f\,\ln_{\kappa}(f)\,]=\lambda
\,\ln_{\kappa}(f/\alpha) \ \ , \label{VII16}
\end{eqnarray}
with
\begin{eqnarray}
&&\alpha=\left(\frac{1-\kappa}{1+\kappa}\right)^{1/2\kappa} \ ,
\label{VII17} \\
&&\lambda=\sqrt{1-\kappa^2} \ . \label{VII18}
\end{eqnarray}

It is important at this point to remark that the function
$\ln^{\kappa}(x)$ does not possess the latter property. For this
reason, we can't use the generalized logarithm and exponential in
the velocity representation, in order to construct a relativistic
statistical theory.

The distribution function  (\ref{VII11}) in the global rest frame
where $U_{\nu}=(c,0,0,0)$ and in absence of external forces i.e.
$A^{\nu}=0$ simplifies as
\begin{eqnarray}
f=\alpha \exp_{\kappa}\left(- \beta\,\frac{W-\mu }{\lambda}\right) \
, \label{VII19}
\end{eqnarray}
being $W$ the relativistic kinetic energy.

We observe that the latter distribution in the classical limit
($\kappa \rightarrow 0$, $W \rightarrow 0$) reduces to the classical
Maxwell-Boltzmann distribution i.e. $f\approx (1/e)\,\exp\left(-
\beta\,(W-\mu) \right)$, while at relativistic energies ($W
\rightarrow +\infty$) presents power law tails $f\propto
W^{-1/\kappa}$, in accordance with the experimental evidence in
several relativistic systems.

\subsection{The H-theorem}

In the ordinary relativistic kinetics it is well known from the
H-theorem that the production of entropy is never negative and in
equilibrium conditions there is no entropy production. In the
following we will demonstrate the H-theorem for the system governed
by the kinetic equation (\ref{VII6}) when the two-particle
correlation function is given by (\ref{VII7}).

By using the property (\ref{VII16}) of $\ln_{\kappa}(f)$ and the
notation $g=f/\alpha$ one obtains
\begin{eqnarray}
\partial_{\nu}(f \ln_{\kappa} f)&&=\left[\frac{\partial}{\partial f}\,f \ln_{\kappa}
f\right]\partial_{\nu}f \nonumber \\ &&=\lambda \ln_{\kappa}
(f/\alpha)\,\partial_{\nu}f \nonumber \\ &&=\lambda \alpha
\ln_{\kappa} (g)\,\partial_{\nu}g \ \ . \label{VII20}
\end{eqnarray}

The entropy production $\partial_{\nu}S^{\nu}$ can be calculated
starting from the definition of $S^{\nu}$ given by Eq.
(\ref{VII12}). By using the result (\ref{VII20}) and the evolution
equation (\ref{VII6}) it obtains the following expression for the
entropy production
\begin{eqnarray}
\partial_{\nu}S^{\nu}= &&-\lambda\,\alpha \int
\frac{d^3p}{p^{0}}\,\ln_{\kappa}(g)\,\,p^{\nu}\,\partial_{\nu}g \nonumber \\
= &&- \lambda \int
\frac{d^3p}{p^{0}}\,\ln_{\kappa}(g)\,\,p^{\nu}\,\partial_{\nu}f \nonumber \\
=&&- \lambda \int
\frac{d^3p'}{{p'}^{0}}\frac{d^3p_1}{p_1^{\,0}}\frac{d^3p'_1}{{p'}_{\!\!1}^{0}}
\frac{d^3p}{p^{0}}\,\,G \nonumber \\
&& \times
\left(g'\otimes\mbox{\raisebox{-2mm}{\hspace{-3.3mm}$\scriptstyle
\kappa$}} \hspace{1mm}g'_1-
g\otimes\mbox{\raisebox{-2mm}{\hspace{-3.3mm}$\scriptstyle
\kappa$}} \hspace{1mm}g_1\,\right) \, \ln_{\kappa}(g) \nonumber \\
&&- \lambda \,m \int
\frac{d^3p}{p^{0}}\,\ln_{\kappa}(g)\,\,F^{\nu}\frac{\partial
f}{\partial p^{\,\nu}}. \ \ \ \label{VII21}
\end{eqnarray}
Since the Lorentz force $F^{\nu}$ has the properties
$p^{\nu}F_{\nu}=0$ and $\partial F^{\nu}/\partial p^{\nu}=0$ the
last term in the above equation involving $F^{\nu}$ is equal to zero
\cite{DEGROOT}, therefore we have
\begin{eqnarray}
\partial_{\nu}S^{\nu}=\!\!\!\!\! &&- \lambda  \int
\frac{d^3p'}{{p'}^{0}}\frac{d^3p_1}{p_1^{\,0}}\frac{d^3p'_1}{{p'}_{\!\!1}^{0}}
\frac{d^3p}{p^{0}}\,\,G \nonumber \\
&& \times
\left(g'\otimes\mbox{\raisebox{-2mm}{\hspace{-3.3mm}$\scriptstyle
\kappa$}} \hspace{1mm}g'_1-
g\otimes\mbox{\raisebox{-2mm}{\hspace{-3.3mm}$\scriptstyle \kappa$}}
\hspace{1mm}g_1\right) \, \ln_{\kappa}(g) . \ \ \ \label{VII22}
\end{eqnarray}
Given the particular symmetry of the integral in the latter equation
we can write the entropy production as follows
\begin{eqnarray}
\partial_{\nu}S^{\nu}=\!\!\!\!\!&&-\frac{1}{4}\, \lambda \int
\frac{d^3p'}{{p'}^{0}}\frac{d^3p_1}{p_1^{\,0}}\frac{d^3p'_1}{{p'}_{\!\!1}^{0}}
\frac{d^3p}{p^{0}}\,\,G \nonumber \\
&&\times
\left(g'\otimes\mbox{\raisebox{-2mm}{\hspace{-3.3mm}$\scriptstyle
\kappa$}} \hspace{1mm}g'_1-
g\otimes\mbox{\raisebox{-2mm}{\hspace{-3.3mm}$\scriptstyle
\kappa$}} \hspace{1mm}g_1\right) \nonumber \\
&&\times \, [\ln_{\kappa}(g) + \ln_{\kappa}(g_1)-\ln_{\kappa}(g') -
\ln_{\kappa}(g'_1) ]\ . \ \ \ \ \ \label{VII23}
\end{eqnarray}
Finally we set the latter equation in the form
\begin{eqnarray}
\partial_{\nu}S^{\nu}=\!\!\!\!\!&&\frac{1}{4}\, \lambda \int
\frac{d^3p'}{{p'}^{0}}\frac{d^3p_1}{p_1^{\,0}}\frac{d^3p'_1}{{p'}_{\!\!1}^{0}}
\frac{d^3p}{p^{0}}\,\,G \nonumber \\
&&\times
\left(g'\otimes\mbox{\raisebox{-2mm}{\hspace{-3.3mm}$\scriptstyle
\kappa$}} \hspace{1mm}\,g'_1-
g\otimes\mbox{\raisebox{-2mm}{\hspace{-3.3mm}$\scriptstyle
\kappa$}} \hspace{1mm}\,g_1\right) \nonumber \\
&&\times \left[\,\ln_{\kappa}\,
(g'\otimes\mbox{\raisebox{-2mm}{\hspace{-3.3mm}$\scriptstyle
\kappa$}} \hspace{1mm}g'_1)-
\ln_{\kappa}\,(g\otimes\mbox{\raisebox{-2mm}{\hspace{-3.3mm}$\scriptstyle
\kappa$}} \hspace{1mm}g_1)\right], \label{VII24}
\end{eqnarray}
and after posing
$h=g\otimes\mbox{\raisebox{-2mm}{\hspace{-3.3mm}$\scriptstyle
\kappa$}} \hspace{1mm}g_1$ and
$h'=g'\otimes\mbox{\raisebox{-2mm}{\hspace{-3.3mm}$\scriptstyle
\kappa$}} \hspace{1mm}g'_1$ and taking into account that
$\ln_{\kappa}h$ is an increasing function, we note that
$(h'-h)\,(\ln_{\kappa}h'-\ln_{\kappa}h)\geq 0$, $\forall \,\,h,h'$.

Consequently we can conclude that
\begin{equation}
\partial_{\nu}S^{\nu}\geq 0 \ \ . \label{VII25}
\end{equation}
This last relation is the local formulation of the relativistic
H-theorem which represents the second law of the thermodynamics for
the system governed by the evolution equation (\ref{VII6}) and the
two-particle correlation function (\ref{VII7}).

\subsection{On the definition of the entropy}

Let us adopt for the two particle correlation function  the
definition
\begin{eqnarray}
C(f,f_1)=
f\,\,\otimes\mbox{\raisebox{-2mm}{\hspace{-3.3mm}$\scriptstyle
\kappa$}} \hspace{1mm}\,\,f_1  \ \ , \label{VII26}
\end{eqnarray}
in place of the (\ref{VII7}). It is straightforward to verify
that one obtains the following expressions for the distribution
function, the four-vector entropy and the scalar entropy
\begin{eqnarray}
&&f=\exp_{\kappa}\bigg(-\beta\,\frac{\left(p^{\nu}+q A^{\nu}\!/c
\right)\,U_{\nu}-mc^2-\mu }{\lambda}\bigg) \ , \ \ \ \ \ \ \ \ \
\label{VII27}
\\
&&S^{\nu}= - \int \frac{d^3p}{p^{0}}\,p^{\nu}\, f
\,\ln_{\kappa}(\alpha f) \ , \label{VII28}
\\
&&S= - \int d^3p \, \, f \,\ln_{\kappa}(\alpha f) \ , \label{VII29}
\end{eqnarray}
in place of (\ref{VII11}), (\ref{VII12}) and (\ref{VII14})
respectively.

In particular, in the classical limits we have $\lim_{\kappa
\rightarrow 0}\alpha=1/e$, and the entropy (\ref{VII29}) reduces to
the classical expression $S_{clas}= - \int d^3p \, \, f \,[\,\ln f
-1 \,]$, used some time in kinetic theory \cite{DEGROOT}.

In order to explain better the differences between the two
definitions (\ref{VII14}) and (\ref{VII29}) of the entropy we
consider a system described by a discrete probability distribution
$f=\{f_i\,, 1 \leq i \leq N\}$. We observe that the probability
distribution $f=\{\delta_{i m}\,, 1 \leq i \leq N \}$ being $m$ a
fixed integer with $1 \leq m \leq N$, describes a state of the
system for which we have the maximum information. It is natural to
set for this state $S=0$. Clearly it is possible only if we adopt
for the system entropy the definition (\ref{VII14}). This
interesting property of the entropy (\ref{VII14}) makes it more
appealing with respect the definition (\ref{VII29}) which predicts a
residual entropy $S=-\ln_{\kappa}\alpha$ for the state corresponding
to the maximum information. Anyway, the two above discussed choices
for the entropy definition does not influence the physics of the
system.

\sect{Conclusions}

We have shown that the special relativity laws and the maximum
entropy principle, suggest a relativistic generalization for the
two-particle correlation function in the relativistic Boltzmann
equation. This fact imply a relativistic generalization  of the
classical Boltzmann-Gibbs-Shannon entropy.

The so obtained, fully relativistic Boltzmann equation, obeys the H-theorem and predicts a stationary stable distribution, presenting power-law high-energy tails, according to the experimental evidence. The ensued relativistic kinetic theory preserves the main features of the classical kinetics which recovers in $c \rightarrow \infty$ limit.

In the last few years the statistical theory based on the new entropy \cite{PhA01,PLA01,PRE02,PRE05}, has been considered by various authors. Investigations
related with the foundations of the theory include e.g. the
H-theorem and the molecular chaos hypothesis
\cite{Silva06A,Silva06B}, the thermodynamic stability
\cite{Wada1,Wada2}, the Lesche stability
\cite{KSPA04,AKSJPA04,Naudts1,Naudts2}, the Legendre structure of
the ensued thermodynamics \cite{ScarfoneWada} etc. On the other hand,
specific applications of the theory, include e.g.
the cosmic rays \cite{PRE02}, relativistic \cite{GuoRelativistic}
and classical \cite{GuoClassic} plasmas  in presence of external
electromagnetic fields, the relaxation in relativistic plasmas under
wave-particle interactions \cite{Lapenta,Lapenta2009}, astrophysical systems \cite{Silva08,Carvalho}, the kinetics
of interacting atoms and photons \cite{Rossani}, particle systems in
external conservative force fields \cite{Silva2008}, the quark-gluon
plasma formation \cite{Tewel} etc. Other applications regard
dynamical systems at the edge of chaos \cite{Corradu,Celikoglu},
fractal systems \cite{Olemskoi}, the random matrix theory
\cite{AbulMagd}, the error theory \cite{WadaSuyari06}, the game
theory \cite{Topsoe}, the Information theory \cite{WadaSuyari07},
etc. Also applications to economic systems have been considered e.g.
to study the personal income distribution
\cite{Clementi,Clementi2008}, to model deterministic heterogeneity
in tastes and product differentiation \cite{Rajaon,Rajaon2008} etc.

%\vfill
%\eject
\end{document}